\documentclass[newstyle,twocolumn,proceedings]{rmaa1}
\usepackage{epsfig}

\def\apjs{ApJS}
\def\apj{ApJ}
\def\mnras{MNRAS}
\def\aj{AJ}
\def\aap{A\&A}
\def\aaps{A\&AS}

\def\apjl{ApJL}
\def\pasp{PASP}

\renewcommand{\P}[1]{%
\ifnum#1=1\hbox{OW~168--326E}\fi
\ifnum#1=2\hbox{OW~167--317}\fi
\ifnum#1=3\hbox{OW~163--317}\fi
\ifnum#1=5\hbox{OW~158--323}\fi
\ifnum#1=0\hbox{OW~171--334}\fi}
\makeatletter
\title{Spectrophotometry of massive star forming regions with GTC}
\author{M. Castellanos\altaffilmark{} and A. I. D\'\i az\altaffilmark{} 
  \affil{Universidad Aut\'onoma de Madrid, Spain}
\and E. Terlevich\altaffilmark{1}\affil{INAOE, M\'exico}}
\altaffiltext{1}{Visiting Fellow, IoA, Cambridge}

\fulladdresses{
\item M. Castellanos and A. I. D\'\i az: Depto. F\'\i sica Te\'orica, C-XI,
  Universidad Aut\'onoma de Madrid, 28049 Madrid,
  Espa\~na (marcelo.castellanos,angeles.diaz@uam.es).
\item E. Terlevich: Instituto Nacional de Astronom\'\i a, Optica y Electr\'onica, 
Tonantzintla, Apdo. Postal 51, 72000 Puebla, M\'exico (eterlevi@inaoep.mx).}

\shortauthor{Castellanos et al.}
\shorttitle{Massive star forming regions with GTC}


\abstract{%
 In these latest years, the detection of Wolf-Rayet (WR) stars in Giant HII Regions (GHRs) has yielded several questions about our current understanding of massive stars evolution and hot expanding atmospheres, the age of the ionizing populations and their impact onto the physical properties of GHRs. Here, we present spectrophotometric observations of 4 extragalactic GHRs which show WR features in their spectra. Our goal is to reproduce simultaneously the observed WR properties and the emission line spectra with the help of current evolutionary synthesis models. Finally, we address the main advantages that GTC will provide to our better understanding of massive star forming regions. }
\resumen{%
 En estos \'ultimos a\~nos, la detecci\'on de estrellas Wolf-Rayet (WR) en regiones HII gigantes ha levantado ciertos interrogantes sobre nuestros conocimientos de la evoluci\'on de estrellas masivas y sus atm\'osferas extensas, la edad de la poblaci\'on ionizante y su impacto sobre las propiedades f\'\i sicas de dichas regiones. En este art\'\i culo, presentamos observaciones espectrofotom\'etricas de 4 regiones HII extragal\'acticas gigantes que muestran rasgos debidos a estrellas WR en sus espectros de emisi\'on. Nuestro objetivo consiste en reproducir de forma simult\'anea tanto las propiedades de las estrellas WR observadas, como el espectro de l\'\i neas de emisi\'on del gas, con la ayuda de modelos de s\'\i ntesis de poblaciones. Finalmente, se\~nalamos las principales ventajas que proporcionar\'a GTC para nuestro mejor entendimiento de regiones de formaci\'on de estrellas masivas. }

\listofauthors{M.~Castellanos, A.~I.~D\'\i az \& E.~Terlevich}
\indexauthor{Castellanos, M.}
\indexauthor{D\'\i az, A.~I.}
\indexauthor{Terlevich, E.}

\begin{document}

\maketitle

\section{Introduction}
\label{sec:intro}
 WR stars are evolved descendants of massive O stars, and hence, they represent
one of the latest stages of massive star evolution. These stars experiment powerful winds, which can lead to a complete loss of their outer envelopes. Let us summarize the main topics associated with the presence of WR stars in GHRs. Stellar evolution predicts mass loss rates to be more prominent at high metallicity (Z $\geq$ Z$_{\odot}$) and, therefore, O stars can enter the WR phase at a lower cutoff mass (Meynet 1995). This leads to both higher WR/O star ratios and stronger recombination lines formed in the wind. However, few detections at high metallicity have been made at the moment (Castellanos, D\'\i az \& Terlevich 2002, Paper I; Bresolin \& Kennicutt 2002) to shed some light on this matter. On the other hand, a high metal content implies an increase in the opacity of the stellar material. This effect would lower the effective temperature of massive stars in regions of high metal content. There is a general agreement about the hardening of the ionising radiation in regions of low metal content (Campbell et al. 1986). However, the softening of the ionising radiation in regions of high metal content is difficult to quantify, since the functional parameters in these regions cannot be parametrised in a trivial way (see D\'\i az et al. 1991). 
Stellar evolution can also predict the WR/O star ratio from the elemental surface abundances and, therefore, it should be able to reproduce adequatelly the WR population properties (mass loss rates, strength of the lines). Again, the detailed observation of GHRs with embedded WR stars is compulsory in order to constrain the stellar evolution assumptions. Finally, taking apart the metal content of the regions, the sole presence of WR stars is supposed to rise up drastically the number of ionising photons at energies higher than $\sim$ 40 eV (P\'erez 1997). Hence, one might expect the ionization structure to change inside GHRs.
In order to analyze these topics, we have studied 4 GHRs in NGC~628 (H13) and NGC~1232 (CDT1, CDT3 and CDT4) showing WR features in their spectra (Castellanos, D\'\i az \& Terlevich 2002, Paper II). 
\section{WR populations} 
\label{sec:WR}
Schaerer \& Vacca (1998; SV98) have presented very detailed models of the WR
population in young star clusters, at different metallicities from
Z=0.001 (1/20 Z$_{\odot}$) to Z=0.04 (2 Z$_{\odot}$). Their clusters are formed
according to a Salperter IMF with upper and lower mass limits of 120
M$_{\odot}$ and 0.8 M$_{\odot}$ respectively. They use the stellar
evolution models by Meynet et al. (1994) which assume a mild
overshooting and  enhanced mass loss rate. These models have been shown
to reproduce the observed WR/O star ratios in a variety of regions
(Maeder \& Meynet 1994). Regarding the energy output of main sequence stars
they use the CoStar models (Schaerer \& de Koter 1997) which include 
non-LTE effects, line
blanketing and stellar winds, for stars with initial masses larger than
20 M$_{\odot}$ and Kurucz (1992) plane-paralel LTE models, including line
blanketing effects, for less massive stars. The atmospheres of evolved
stars in the WR phase correspond to the spherically expanding, non-LTE,
unblanketed models by Schmutz, Leitherer \& Gruenwald (1992).   

SV98 provide accurate predictions, as a function of the
cluster age, for the total number of WR stars and their subtype
distribution, the broad stellar emission lines and the luminosities and
equivalent widths of the two ``WR bumps'' at $\lambda$ 4650 and $\lambda$ 5808 {\AA} . We 
have used these models to derive the age of the ionising population which
contains WR stars and whose metallicity has been previously derived by means of the calculated ionic temperatures (see Paper I). 

Figure 1 shows the predicted emission line intensities and equivalent widths of the WR blue ``bump'' for different metallicities (0.2Z$_{\odot}$; 0.4Z$_{\odot}$ and Z$_{\odot}$) together with the observed values in the four regions. Solid symbols correspond to the observed data. It can be seen that in three of the analyzed regions, observations and predictions are in excellent agreement for a single instantaneous burst. The derived ages for the WR population range between 3.1 Myr (CDT3) and 4.1 Myr (H13 and CDT4). In fact, for regions CDT3 and CDT4
observations and predictions are nearly identical for the respective
clusters and therefore only one symbol is shown. In the case of region CDT1, two open symbols are shown, one corresponding to
the predictions for a single ionising cluster of 2.4 Myr, and another one
including the contribution to the H$\beta$  and continuum luminosity of a cluster not containing WR stars (around 7 Myr). This latter one is able to reproduce the observed WR features and the H$\beta$ equivalent width.


\begin{figure*}
\setcounter{figure}{1}
 \psfig{figure=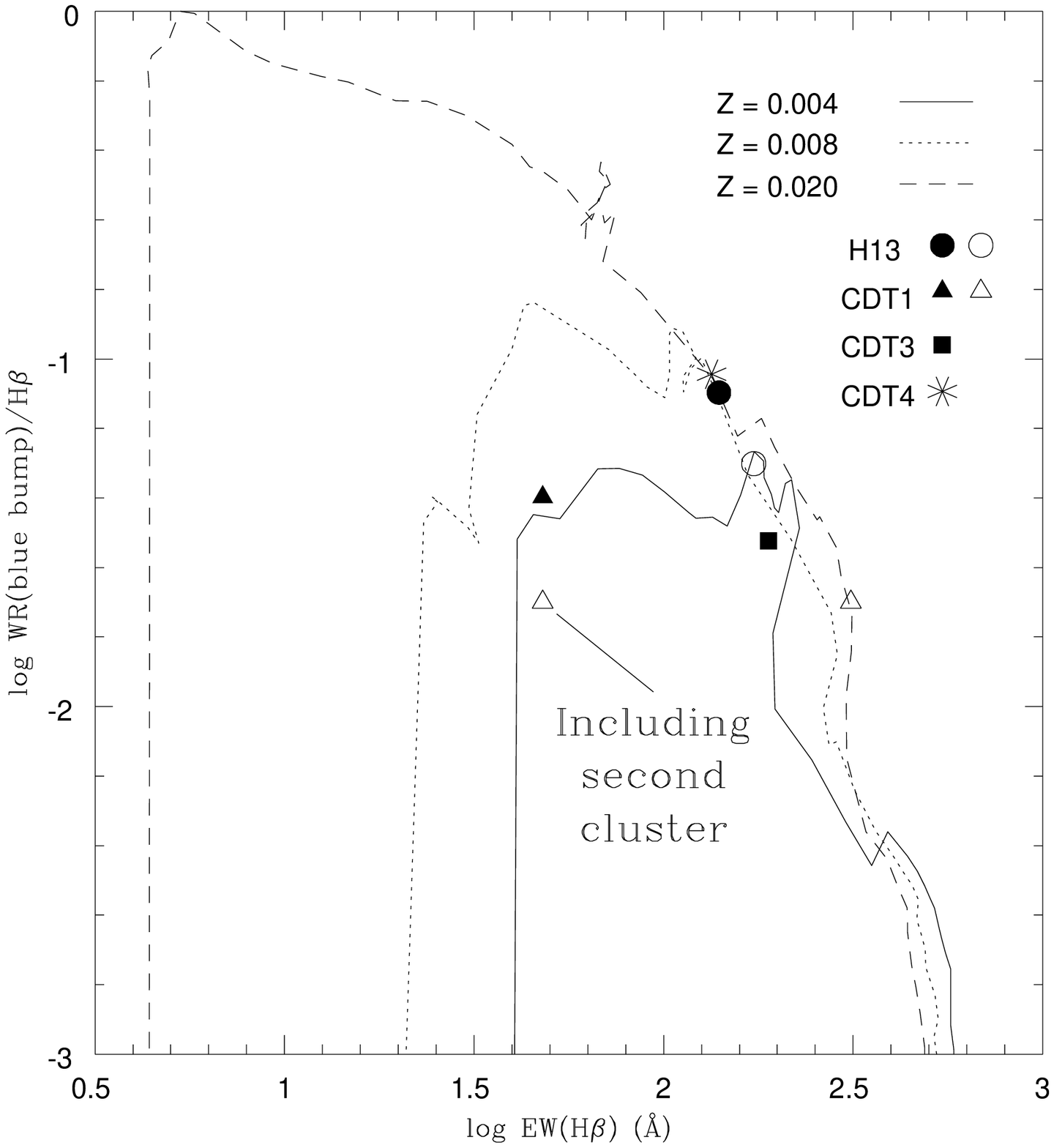,height=6.0cm,width=7.5cm,clip=}
 \psfig{figure=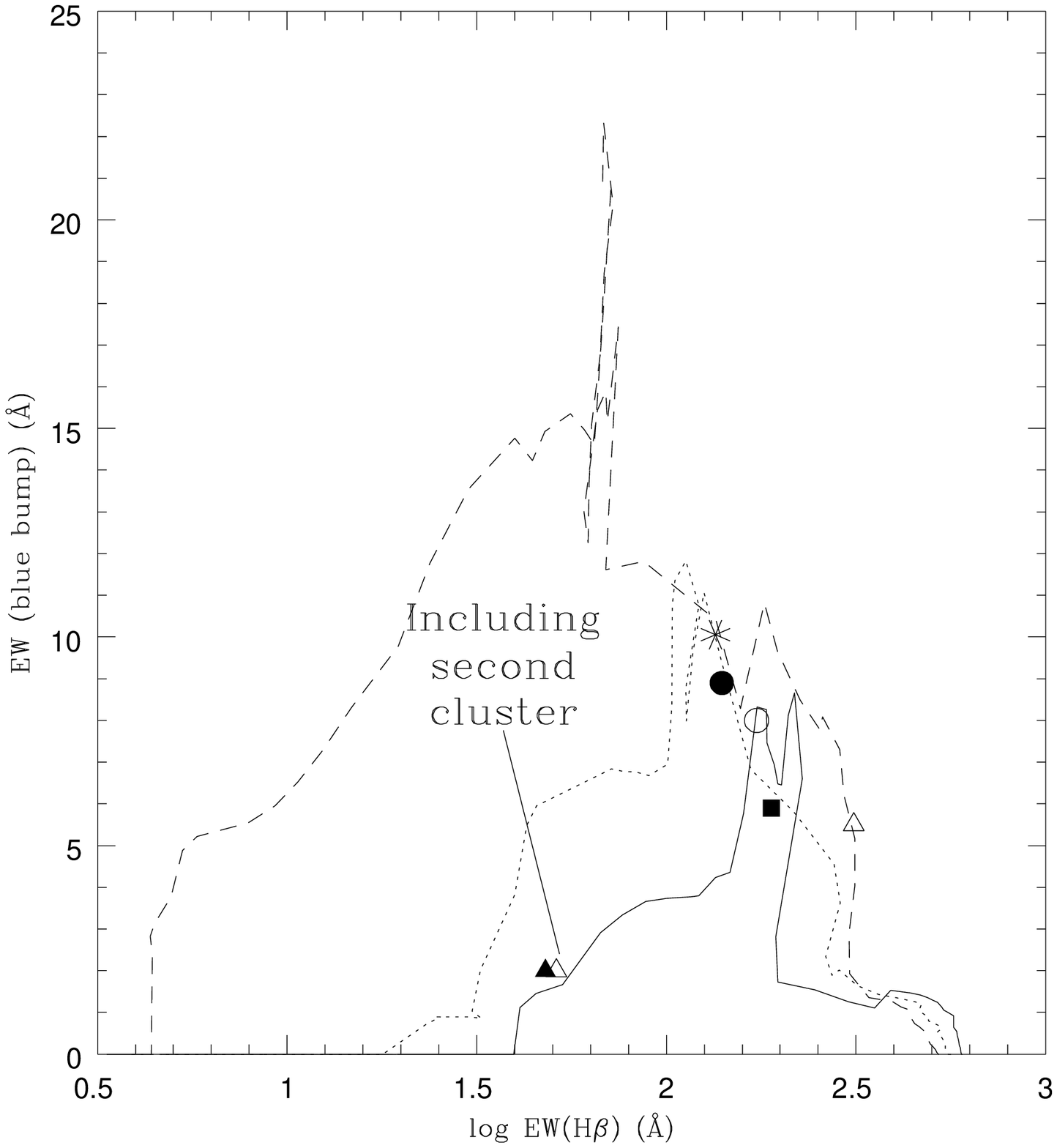,height=6.0cm,width=7.5cm,clip=}
 \caption{Relative intensity (left) and equivalent width (right) of the
   WR blue `bump' {\it versus}  H$\beta$ equivalent width for SV98 models of
   three different metallicities as labelled. The data are shown as
   solid and open symbols as explained in the text.}
\end{figure*}


\section{Modelling the emission line spectra}
\label{sec:mod}
Regarding the integrated Spectral Energy Distribution (SED) from the 
ionising cluster, the recent
models from Leitherer et al. (1999; STARBURST99) provide an almost
selfconsistent frame to be used in combination with the WR models
described above. In fact, they use the same stellar evolution models
with enhanced mass loss rate, the same atmosphere models to describe
the stars in the WR phase and cover the range of metallicities and IMF
used by SV98. The only appreciable difference between both
sets of models concerns the atmospheres of the main sequence stars with
initial masses greater than 20 M$_{\odot}$ which in STARBURST99 are
represented by the plane-paralel Kurucz's models implemented by 
Lejeune, Cuisiner \& Buser (1997). We have
therefore used the STARBUST99 models in order to fit the emission line
spectra for our analysed regions. 
Hence, as the WR emission features have been used to date the ionizing star
clusters, these should reproduce the observed gas emission line 
intensities if the other parameters controlling the emission line
spectra, namely elemental abundances, particle density and ionisation
parameter, are known. Our four observed GHRs meet all these 
requirements.

Therefore, since both SV98 and STARBURST99 models make
use of the same stellar evolution prescriptions, we have
assumed that a STARBURST99 model of a cluster of a given metallicity
which contains the same WR/O star ratio as SV98 (corresponding to a given age), 
must provide the spectral energy 
distribution of the ionising radiation. In the case of region H13 in NGC 628, this model (corresponding to an age of 4.1 Myr from the previous analysis of the WR features)  
however provides an
[OIII]$\lambda$5007 {\AA} line emission which is higher than observed by a factor
of about 4. On the other hand, our best photoionisation
model corresponds to an age of 4.7 Myr producing an H$\beta$
equivalent width of 108 {\AA}, to be compared with the observed one of 140 {\AA}. The rest
of the spectrum is reproduced remarkably well taking into account the uncertainties in both 
the observed values and model computations. We can
conclude that a single star cluster with age between 4.0 and 4.7 Myr
fit all the observations adequately.

For region CDT1, given that a young 2.4 Myr cluster is successful 
in reproducing both the WR blue `bump' luminosity and equivalent 
width (although overpredicts the equivalent width of H$\beta$ 
by a factor of seven), a composite population with, at least 
two clusters must be invoked. We have therefore run a 
model using as ionising source the combination
of the spectral energy distributions 
of two ionising clusters of 2.4 and 7.1 Myr of age calculated with the
STARBURST99 code in which the youngest of
the two provides 10 times  the number of ionising photons emitted by
the oldest. This cluster contains the same WR/O  star ratio as
the SV98 model reproducing the observed WR features. The computed
photoionisation 
model is able to reproduce all the observables adequately.

For the other two GHRs in NGC 1232, CDT3 and CDT4, STARBURST99 
model clusters with the same WR/O ratio as those given by SV98, provide
[OIII]$\lambda$5007 {\AA} line
emission higher than observed by factors of about 5 and 3 respectively.
Clusters slightly older, also
reproducing the observed WR features within the errors, have larger
WR/O number star ratios (up to 8.3 $\times$10$^{-2}$) and therefore their
spectral energy distributions result even harder. Alternatively, it is possible 
to find a combination of ionising clusters which reproduces well the emission
line spectrum, but predicts WR feature luminosities and equivalent
widths well below the observed values.

\section{Discussion and Conclusions}
\label{sec:dis}
The fact that the WR features are adequately reproduced by SV98 models
seem to imply that the evolutionary tracks are able to predict the
right relative numbers of WR/O stars, and their different subtypes at the
derived abundances. These relative numbers, combined with the observed 
emission line luminosities of the individual WR stars, and the predicted
continuum energy distribution of the ionising population predict
emission line intensities and equivalent widths of the WR stars that are in
excellent agreement with observations.
On the other hand, in two of the analyzed regions (CDT3 and CDT4 in NGC 1232), it is
not possible to fit the emission line spectrum since the
population containing WR stars produces a spectral energy distribution
which is too hard to explain the emission of the gas. The same sort of
effect has been found by D\'\i az et al. (2000a) for region 74C in NGC~4258
and Esteban et al. (1993) for the galactic WR
nebula M1-67. These latter authors, from stellar and nebular
spectroscopic analyses,
concluded that lower temperatures were required from the
photoionisation models for late type WN (WNL) stars. In their study
they used the unblanketed WR models of Schmutz et al. (1992). A
subsequent reanalysis of this region has been made by Crowther et al.
(1999) using
blanketed model atmospheres. In this case the resulting ionising
spectrum is much softer and reproduces better the observations although
some discrepancies still remain.
For the high metallicity GHR CDT1 in NGC 1232, a composite population can explain
adequately both the WR features and the emission line spectrum. Composite 
populations for HII regions have been found in previous
works by Mayya \& Prabhu (1996) and by D{\'\i}az et al. (2000b) for disc and
circumnuclear objects respectively from broad band and H$\alpha$ photometry.

In the case of region H13 in NGC 628, a single instantaneous burst between 4.0 and
4.7 Myr, is able to reproduce all the observables within the errors. This result seems
to indicate that line blanketing effects at low metallicity (0.2 Z$_{\odot}$) could 
be less severe for the correct interpretation of the emission line spectra.

Our observations indicate that no appreciable change is found in the ionization
structure of the analyzed HII regions, despite the presence of WR stars. Line blanketing
in WR atmospheres would point again in the right direction. It should be kept in mind that
all our analysis is
based on ionisation
bounded models. An approach of the emission line spectra of
HII region with WR features on the basis of matter bounded models has
been presented in Castellanos, D{\'\i}az \& Tenorio-Tagle (2002). Models of 
this kind have
been found to provide excellent fittings to the observations of H13 in
NGC~628, CDT3 in NGC~1232 and 74C in NGC~4258 using both SV98 and
STARBURST99 models. These models would point to an important leakage
of ionising photons depending on both the metallicity and evolutionary
state of the region.
There are still several caveats about our understanding of emission line
spectra of GHRs with WR stars. Certainly, the controversial observation of
main sequence hydrogen-rich WNL-like stars (de Koter, Heap \& Hubeny 1997; Crowther
\& Dessart 1998), not predicted by current stellar evolution calculations at ages less
than 2Myr at low metallicity, and the observation of dust envelopes around WC stars (Marchenko
 et al. 2002) may have also important consequences to interpret correctly the ionization
structure in GHRs, closely related to the impact of WR stars in the surrounding medium.
The combination of GTC + OSIRIS spectroscopy would allow the observation of WR features in low
excitation HII regions (high metallicity regions, in general) at shorter exposure times. The
detection of He II at 4686{\AA} with Tunable Filters seems to be a promising tool with GTC.
Clearly, GTC would allow to solve some of the problems mentioned above.

\adjustfinalcols

\end{document}